\documentclass[twocolumn,aps,showpacs,preprintnumbers,amsmath,amssymb]{revtex4}

\usepackage{color}
\definecolor{Violet}{rgb}{0,0,0.6}

\usepackage[linkcolor=Violet, colorlinks=true, citecolor=Violet, bookmarks=false]{hyperref}

\usepackage{graphicx,rotating}  
\usepackage{amsmath,amssymb}
\usepackage{dcolumn}
\usepackage{bm}
\newcommand{\unit}[1]{\ensuremath{\,\mathrm{#1}}}

\newcommand{\Hz}{\unit{Hz}\hspace{1mm}}
\newcommand{\MHz}{\unit{MHz}\hspace{1mm}}

\begin{document}

\title{Quantum Simulation of Spin Models on an Arbitrary Lattice with Trapped Ions}
\author{S. Korenblit$^1$, D. Kafri$^1$, W. C. Campbell$^1$, R. Islam$^1$, E. E. Edwards$^1$, 
Z.-X. Gong$^2$, G.-D. Lin$^2$, L.-M. Duan$^2$, J. Kim$^3$, K. Kim$^4$, and C. Monroe$^1$}
\affiliation{$^1$ Joint Quantum Institute, University of Maryland Department of Physics and
                  National Institute of Standards and Technology, College Park, MD  20742 \\
             $^2$ Department of Physics and MCTP, University of Michigan, Ann Arbor, MI 48109 \\
             $^3$ Fitzpatrick Institute for Photonics, Department of Electrical and 
                  Computer Engineering, Duke University, Durham, NC  27708 \\
             $^4$ Center for Quantum Information, Tsinghua University, Beijing 100084, China}
\date{\today}
\begin{abstract}
A collection of trapped atomic ions represents one of the most attractive platforms for the 
quantum simulation of interacting spin networks and quantum magnetism.  
Spin-dependent optical dipole forces applied to an ion crystal create long-range effective spin-spin interactions and 
allow the simulation of spin Hamiltonians that possess nontrivial phases and dynamics.  
Here we show how appropriate design of laser fields can provide for arbitrary multidimensional spin-spin interaction graphs even for the case of a linear spatial array of ions.
This scheme uses currently existing trap technology and is scalable to levels where
classical methods of simulation are intractable.
\end{abstract}
\maketitle


Quantum systems are notoriously difficult to model efficiently using classical computers, owing to the 
exponential complexity in describing a general quantum state as the system grows in size.  
In the 1980s, Richard Feynman proposed to circumvent this problem by employing a control quantum system with tailored interactions 
and logic gates between quantum bits (qubits) in order to simulate the quantum system under investigation \cite{Feynman82, Lloyd96}.

Quantum spin models such as the Ising model have become a proving ground for Feynman's proposal, with systems of qubits behaving 
as effective spins and interactions engineered with external electromagnetic fields. 
The quantum Ising model is the simplest spin Hamiltonian that exhibits nontrivial aspects 
of quantum magnetism such as spin frustration, phase transitions \cite{Sachdev}, and poorly understood 
spin glass and spin liquid phases \cite{SpinGlass,SpinLiq}.  
Indeed, solving for the ground state configuration of spins subject to a general fully-connected Ising
interaction is known to be an NP-complete problem \cite{NP}.  Here we show how a fully-connected Ising or more general Heisenberg spin model with arbitrary couplings across the spin network can be generated in a scalable system of trapped atomic ions, even for a one-dimensional chain in space.  
This may allow quantum simulations with hundreds of spins, where the physics cannot generally be predicted otherwise.

Isolated atoms are an ideal control quantum system for quantum simulations, 
as they exhibit very long coherence times, can be measured with high efficiency, and
can obviously be replicated with nearly identical characteristics.
There has been great progress in the use of atoms trapped in optical lattices for studies of transport phenomena 
such as those described by Hubbard models \cite{Bloch08} and the fractional quantum Hall effect \cite{Lin11}. 
Following original proposals to exploit atomic ions for the simulation of spin models \cite{Porras04, Deng05, Taylor08},
there has been steady progress in simulating Ising models with an effective transverse magnetic field,
showing the behavior of quantum phase transitions in the limit of less than ten spins 
\cite{Schaetz08, Kim09, Kim10, Edwards10, InnsbruuckSim10, Islam11, Lanyon11}.  Recently, the Ising model with an axial
field was demonstrated in neutral atoms in optical lattices with nearest-neighbor couplings \cite{Simon11}. 
All of these studies involve global spin-dependent optical dipole forces to 
generate trivial forms of the spin couplings, such as a uniform
ferromagnet.  Below we show how to tailor optical forces to generate arbitrary 
fully-connected networks of $N$ spins that uniquely specify each of the $N(N-1)/2$ pairwise interactions.  
The scheme is independent of the spatial geometry of the ion crystal and is compatible with 
one-dimensional arrays of trapped ions used in current experiments.

We start with the arbitrary fully-connected Ising Hamiltonian on $N$ spins,
\begin{equation}
\label{effective}
	H = \sum_{i < j} J_{i,j} \sigma_x^{(i)}\sigma_x^{(j)} ,
\end{equation}
where the matrix $J_{i,j}$ is the strength of the spin-spin coupling between atoms $i$ and $j$. 
The Pauli spin operator $\sigma_x^{(i)}$ refers to the effective spin-$1/2$ system within each atom, 
here represented for example by a pair of hyperfine ground states separated by frequency $\omega_{s}$ \cite{YbQubit}.

The spins are coherently manipulated through a pair of counter-propagating laser beams that drive stimulated Raman transitions 
between the spin states while also coupling off-resonantly to the collective motion of the atomic chain \cite{NIST,WinelandBlatt08}.
The atoms are arranged in a linear array, as is typical in a linear radiofrequency ion trap, although this scheme can also apply to other spatial geometries \cite{Taylor08,Clark09, DidiSim09}. 
When the difference frequency between the Raman fields is bichromatic, 
with two spectral components tuned symmetrically at $\omega_{s} \pm \mu$ with $\mu \ll \omega_{s}$, 
the effective spin-spin interaction of Eq. \ref{effective} emerges, mediated by the Coulomb-coupled motion of 
the atomic ions crystal \cite{MS, MSGen}. The axis of the spin-spin interaction on the Bloch sphere can be controlled by the relative optical phase of the two Raman fields \cite{Haljan1ion,Lee05}.
We assume the Raman lasers have wave vector difference $\delta k$ along the principal $X-$axis of transverse motion 
of the ion crystal \cite{TransverseModes} so that each $X-$transverse normal mode $m$ 
with frequency $\omega_m$ contributes to each spin-spin coupling \cite{Kim09},
\begin{equation}
J_{i,j}=\Omega_i\Omega_j\sum_{m=1}^{N}{\frac{\eta_{i,m}\eta_{j,m}\omega_m}{\mu^2-\omega_m^2}}.
\label{coupling}
\end{equation}
Here, the Lamb-Dicke parameter $\eta_{i,m}=b_{i,m}\delta k \sqrt{\hbar / (2 M \omega_m)}$ sets the scale for the 
coupling between spin $i$ and mode $m$, $b_{i,m}$ is the normal mode transformation matrix, 
$M$ is the mass of a single atom, and the single spin Rabi frequency $\Omega_i$ is proportional to the optical intensity at atom $i$.
The above expression is valid when the transverse modes are all laser-cooled to the Lamb-Dicke limit \cite{NIST,DidiRMP},
which is routine in typical linear traps \cite{Kim09}.
We also assume the symmetric detuning $\mu$ is set sufficiently far from all motional sidebands 
($\left|\omega_m-\mu\right|>>\eta_{i,m}\Omega_i$), so that the phonon states can be adiabatically eliminated, leaving the pure spin-spin coupling above \cite{MS,MSGen,Kim09}.

\begin{figure}
\includegraphics[width=0.99\linewidth]{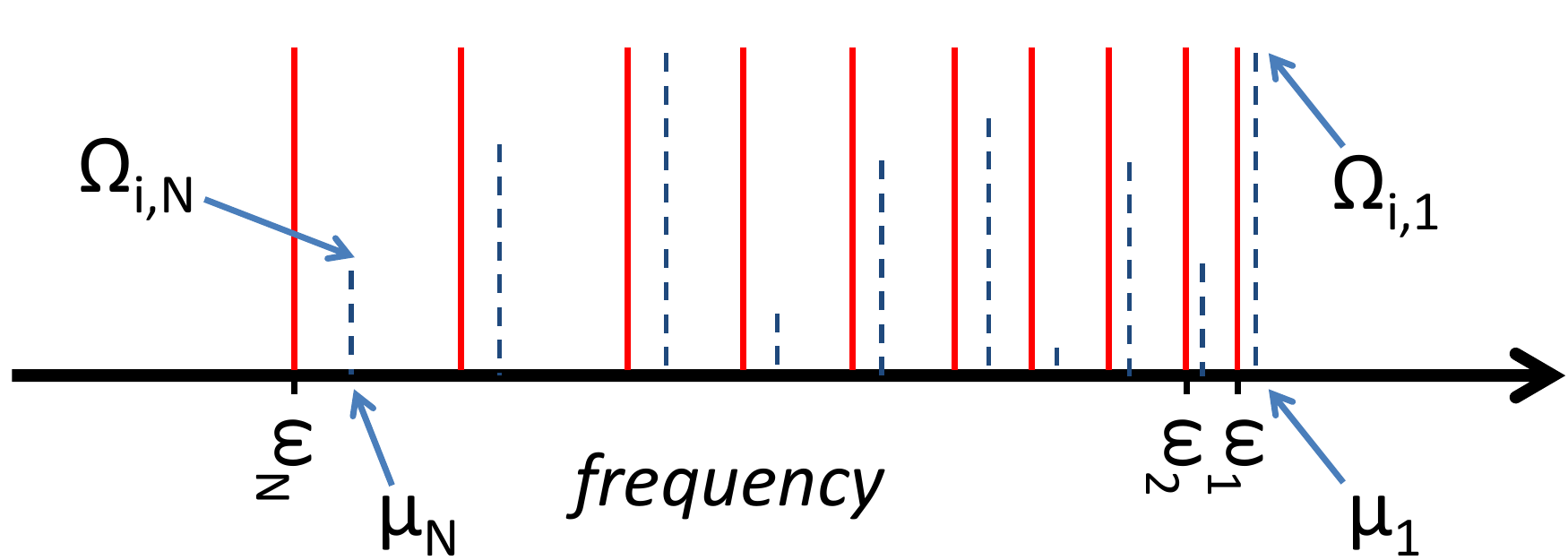}
\caption{
Spectrum of transverse mode frequencies $\omega_m$ for $N=10$ ions in an anisotropic harmonic linear trap (solid lines), with the highest mode frequency $\omega_1$ corresponding to center-of-mass motion.  Raman beatnote detunings $\mu_m$ from the qubit frequency $\omega_{s}$ are denoted by the $N=10$ dashed lines, with each spectral feature near a given motional sideband. The height of the dashed lines represents the intensity of each beatnote for ion i. In general each ion will be illuminated with a different set of intensities.}
\label{ions}
\end{figure}

The above expression has $N+1$ control parameters in the set of Rabi frequencies $\{ \Omega_i \}$ 
and the global beatnote detuning $\mu$.  In order to generate an arbitrary Ising coupling matrix $J_{i,j}$ however,
it is necessary to have at least $N(N-1)/2$ independent controls.
Additional control parameters can be introduced by adding multiple spectral beatnote detunings to the Raman beams, 
one near each motional mode (see Fig. \ref{ions}), with a unique pattern of spectral components on each ion.
There are several ways to achieve this, all involving some form of individual ion addressing.
For simplicity, we retain the same set of $N$ Raman beatnote detunings $\{ \mu_m \}$ on each ion and 
allow the spectral amplitude pattern to vary between ions, all characterized by the $N \times N$ Rabi frequency matrix $\Omega_{i,n}$ 
of spectral component $n$ at ion $i$. Note that the relative signs of the Rabi frequency matrix elements can be controlled by 
adjusting the phase of each spectral component. This individual spectral amplitude addressing provides $N^2$ control parameters, 
and the general Ising coupling matrix becomes
\begin{eqnarray}
J_{i,j} &=& \sum_{n=1}^{N}\Omega_{i,n}\Omega_{j,n}\sum_{m=1}^{N}{\frac{\eta_{i,m}\eta_{j,m}\omega_m}{\mu_n^2-\omega_m^2}} \label{J1} \\
        &\equiv& \sum_{n=1}^{N}\Omega_{i,n}\Omega_{j,n} F_{i,j,n},
\label{J}
\end{eqnarray}
where $F_{i,j,n}$ characterizes the response of Ising coupling $J_{i,j}$ to spectral component $n$.  
An exact derivation of the effective Hamiltonian given a spectrum of spin-dependent forces gives rise to new off-resonant cross terms, which can be shown to be negligible in the rotating wave approximation, as long as the detunings are chosen so that their sums and differences do not directly encroach any sideband features in the motional spectrum of the crystal \cite{TransverseModes}.



We tune each beat note frequency near a unique normal mode so that $F_{i,j,n}$ has independent contributions for each $n$.
Given a desired Ising coupling matrix $J_{i,j}$, we use standard constrained nonlinear optimization to find the corresponding Rabi frequency matrix $\Omega_{i,n}$, while minimizing the total beam intensity. The deviation between the desired and the attained coupling was less than typical round-off errors.

\begin{figure}
\raggedright
\hspace{0.3\linewidth}(a)\hspace{0.41\linewidth}(b)
\includegraphics[trim =0.1\linewidth 10mm 16mm 15mm 42mm, clip, 
width=0.6\linewidth]{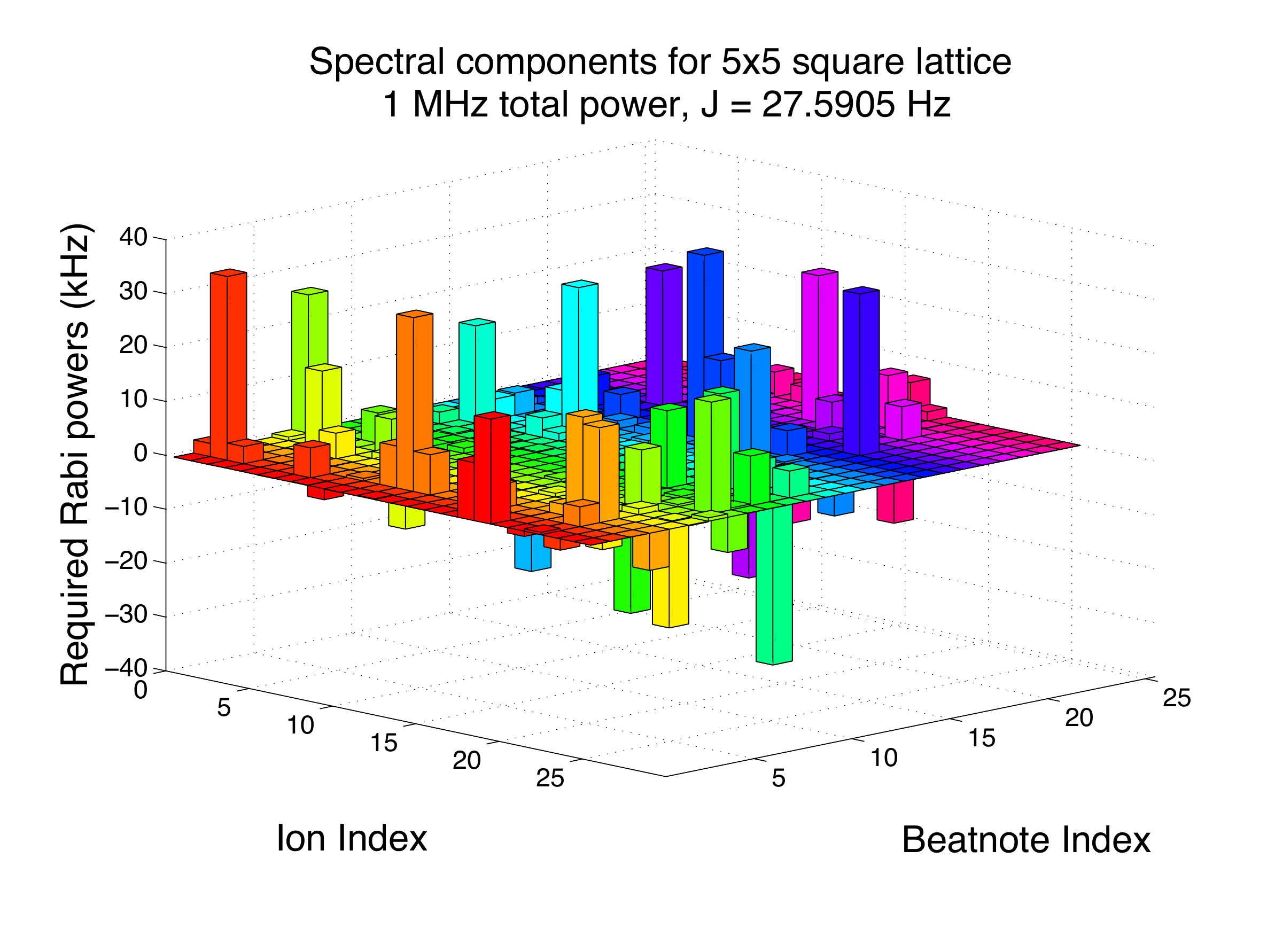}\includegraphics[width=0.38\linewidth]{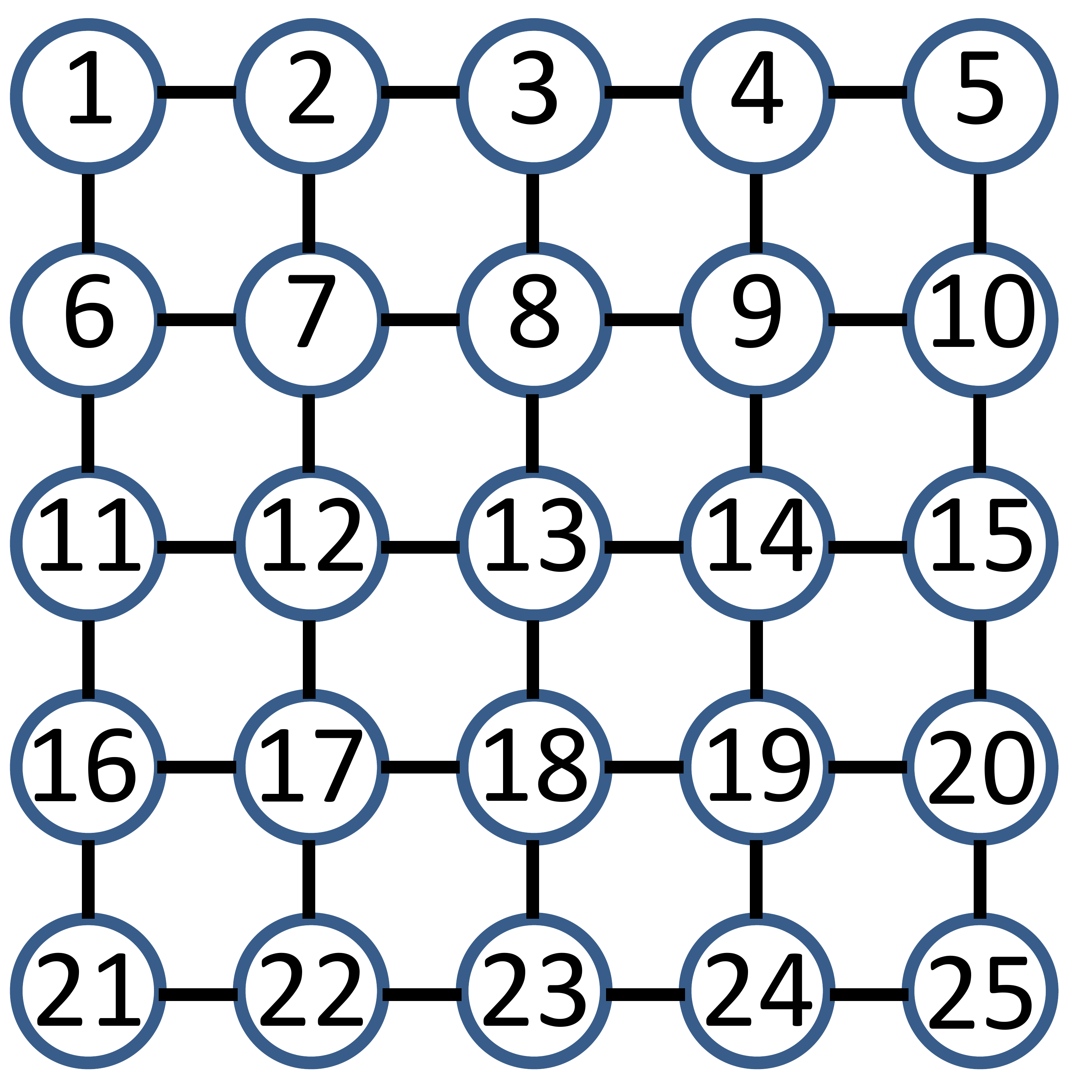}\\
\hspace{0.5\linewidth}(c)
\includegraphics[width=0.99\linewidth]{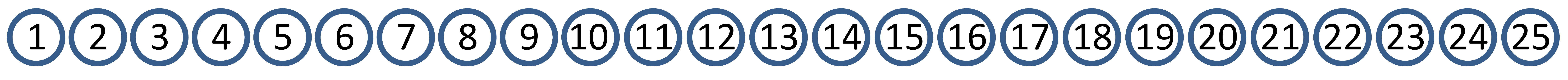}\\
\hspace{0.3\linewidth}(d)\hspace{0.41\linewidth}(e)
\includegraphics[trim =0.1\linewidth 10mm 18mm 23mm 43mm, clip, width=0.6\linewidth]{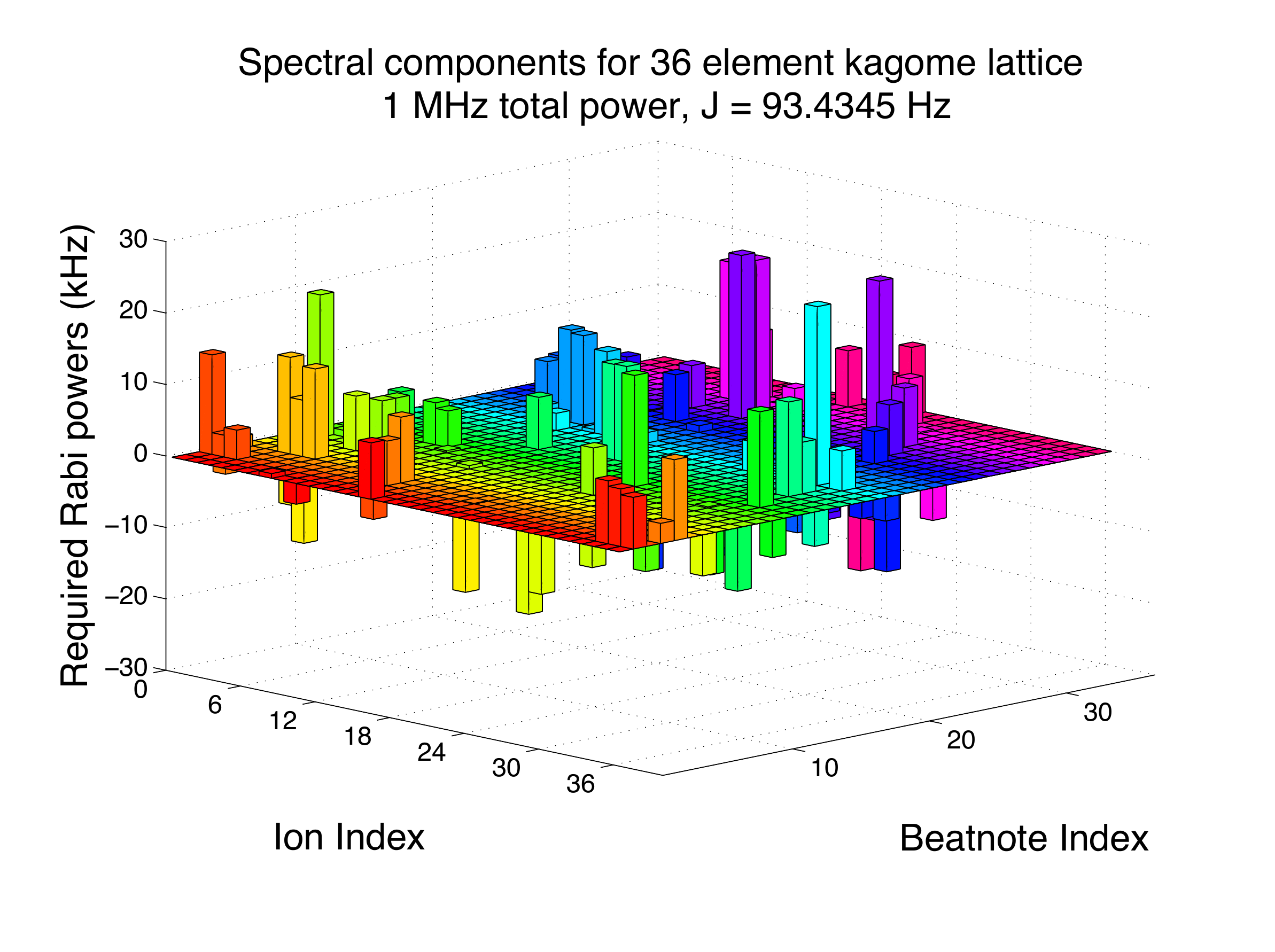}\includegraphics[width=0.38\linewidth]{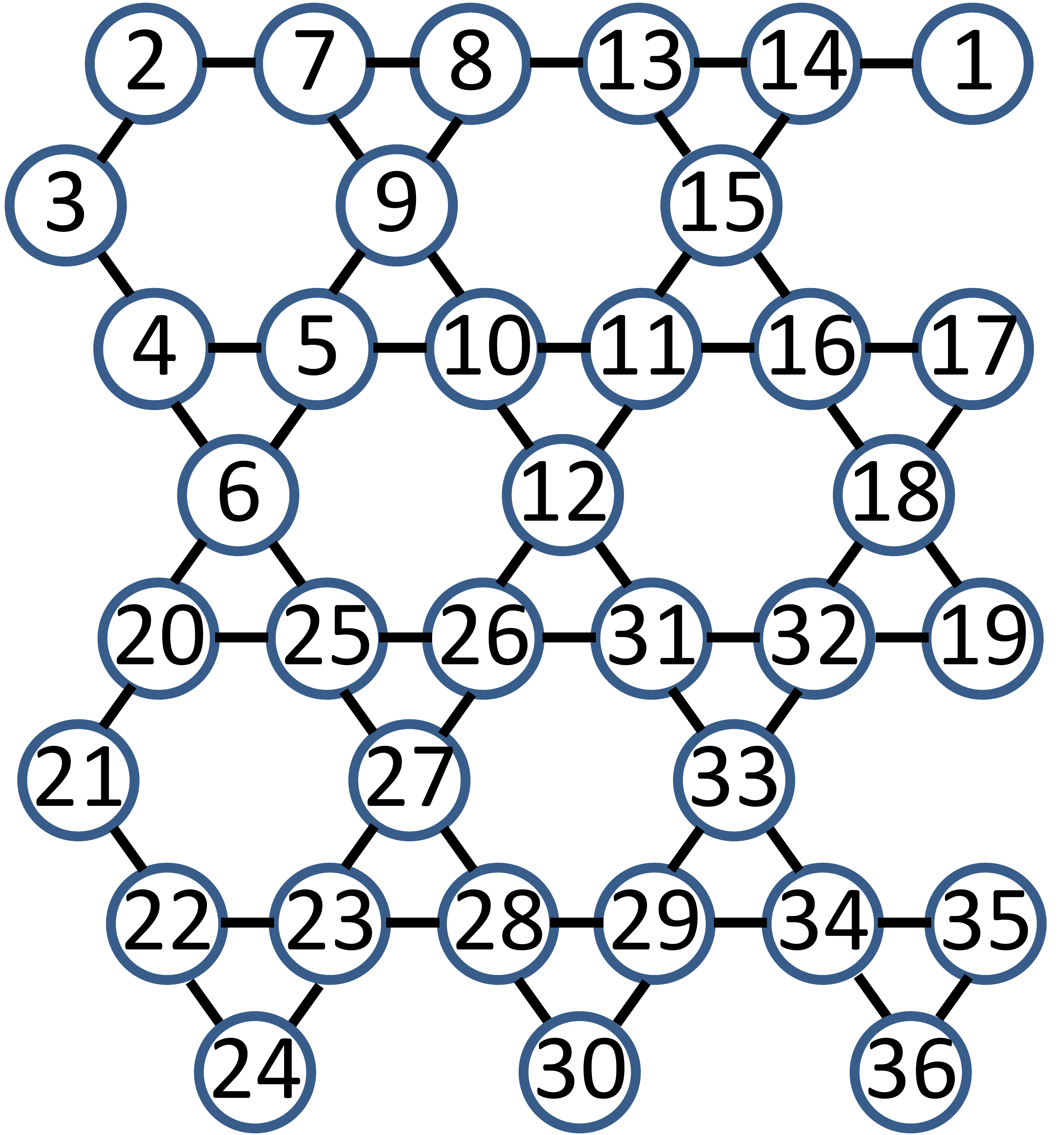}

\caption{(a) Calculated Rabi frequency matrix $\Omega_{i,n}$ to generate $2$D square lattice shown in (b), using the linear chain of $N=25$ ions shown in (c). The ion index refers to the order in the linear chain.  The attained $J_{i,j}$ nearest-neighbor is $27.6$ \Hz for $f_s=0.1$. (d) Calculated Rabi frequency matrix $\Omega_{i,n}$ to generate $2$D Kagome lattice shown in (e) using a linear chain of $N=36$ ions. The attained $J_{i,j}$ nearest-neighbor is $93.4$ \Hz for $f_s=0.03$. In both cases the total optical intensity corresponds to a Rabi frequency of $1$ \MHz if focused on a single ion, the nearest-neighbor couplings are antiferromagnetic and we impose periodic boundary conditions.}
\label{fig:graph}
\end{figure}

We now present two example solutions for $\Omega_{i,n}$ that produce interesting interaction graph topologies. 
First we calculate a Rabi frequency matrix that results in a $2$D square lattice of nearest-neighbor antiferromagnetic couplings with $N=25$ ions ($5 \times 5$ grid with periodic boundary conditions), shown in Figs. \ref{fig:graph}a-b.
Next we produce a $2$D Kagome lattice of antiferromagnetic interactions, a geometry that 
can support high levels of geometrical frustration \cite{frust}, shown in Figs. \ref{fig:graph}d-e. 
In both cases we assume the center-of-mass (COM) mode to be $\omega_1/2\pi = 5$ \MHz, and a fixed total optical intensity corresponding to $\sum_{i,n}|\Omega_{i,n}|=1$ \MHz.
The beatnote frequencies $\mu_m$ are each tuned blue of the mode $m$ sideband by a fraction $f_s$ of the spacing $\omega_1 - \omega_2$ between the most closely-spaced modes (the COM and ``tilt" modes, see Fig. \ref{ions}), which itself scales as $\text{log}N/N^2$.
In these examples, the sparse nearest-neighbor nature of the interaction graphs require that most of the Ising interactions vanish, indicating a high level of coherent control over all of the Ising couplings.  

\begin{figure}
\includegraphics[trim = 0mm 20mm 30mm 2mm, clip, width=0.99\linewidth]{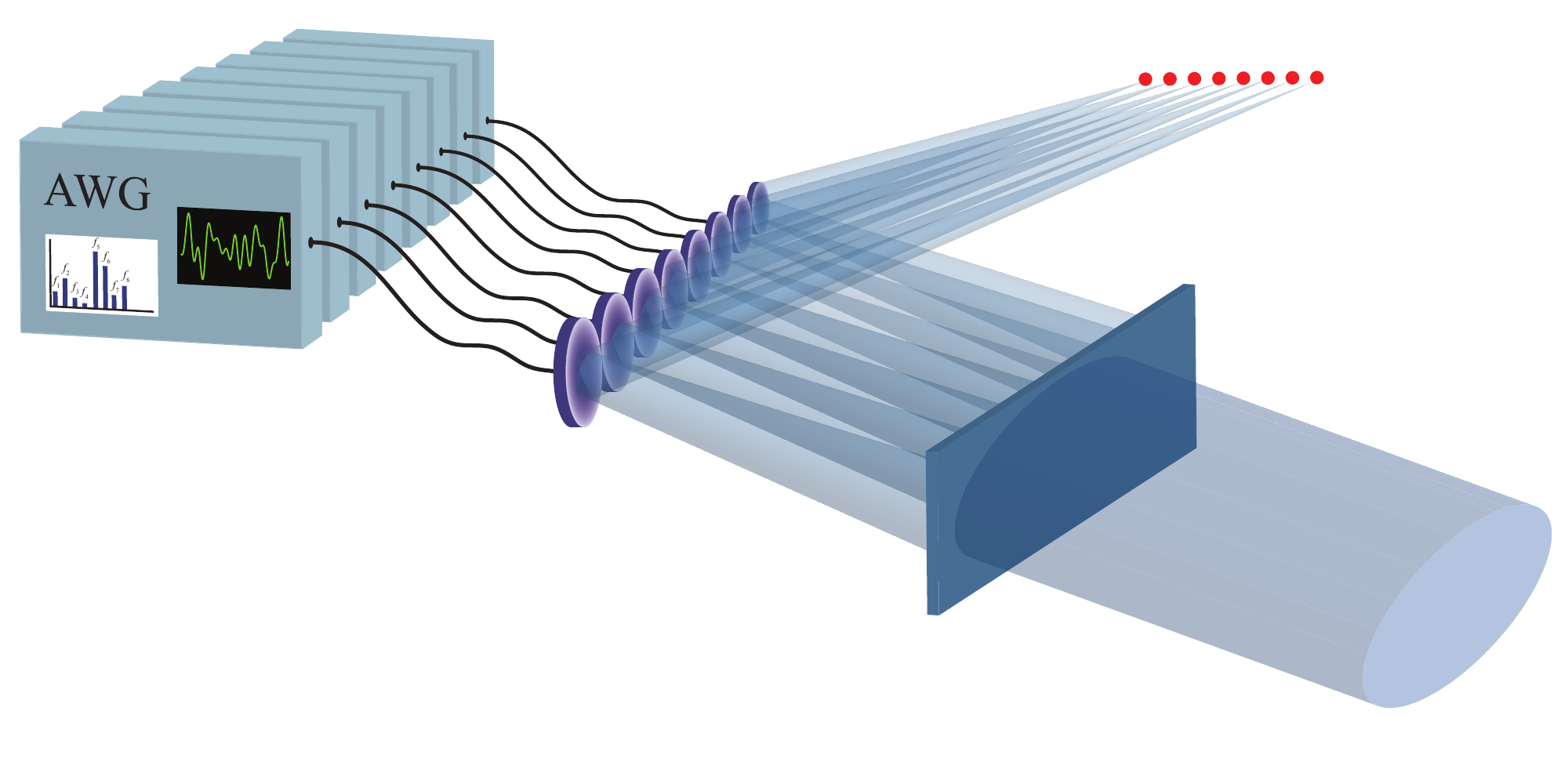}
\caption{
Schematic for individual spectral addressing a linear chain of $N$ ions.  A laser beam is split into a linear array of spots that each traverse $N$ independent acousto- or electro-optical modulators, driven by an $N$ independent arbitrary waveform generators (AWG).  Alternatively, as discussed in the text, the beam can be broken into an array of $N^2$ beams that strike a $N \times N$ array of micromirrors each independently modulated, or a spatial light modulator.}
\label{fig:schematic}
\end{figure}

In order to generate a unique spectrum of Raman beams for each of $N$ ions, some type of individual addressing is necessary.
For simplicity, we assume one of the two Raman beams is uniform and monochromatic, and the high frequency beatnote near 
the qubit frequency $\omega_{s}$ or other global offset frequencies can be set by tuning this monochromatic beam. 
We focus attention on providing the requisite frequencies of the second beam, spread over 
a range given by the bandwidth of the transverse motional mode frequencies of the ion chain, typically in the range $1-5$ \MHz. 
We suggest three possible methods for providing spatial dependent frequency modulation to one of the Raman beams.
The first method (Fig. \ref{fig:schematic}) splits a single beam with a linear chain of $N$ individual optical modulators (e.g., acoustooptic or electrooptic devices), driven by $N$ independent arbitrary waveform generators.
The second method splits a single monochromatic beam into a $N \times N$ square grid 
and directs them onto a $2$D array of $N^2$ micromirrors \cite{JKim10} that are each individually phase modulated at a single frequency (and phase) \cite{JKim11} and finally focused on the ion chain.  
The third method again splits the beam into an $N \times N$ grid of beams, this time with the vertical direction split by a single
acoustooptic modulator, correlating beam position to frequency.  This beam is then directed into a spatial light modulator 
that acts to mask (or phase shift) each of the $N \times N$ beams independently, and again focused onto the ion chain.
In these implementations, it may be desirable to work with a uniformly spaced array of ions in the linear trap, so that 
the modulating elements are also uniformly spaced.  This can be accomplished by using a quartic or higher order linear trap 
\cite{Anharmonic,GTRIlin}.

\begin{figure}
\includegraphics[trim = 16mm 6mm 17mm 10mm, clip, width=0.99\linewidth]{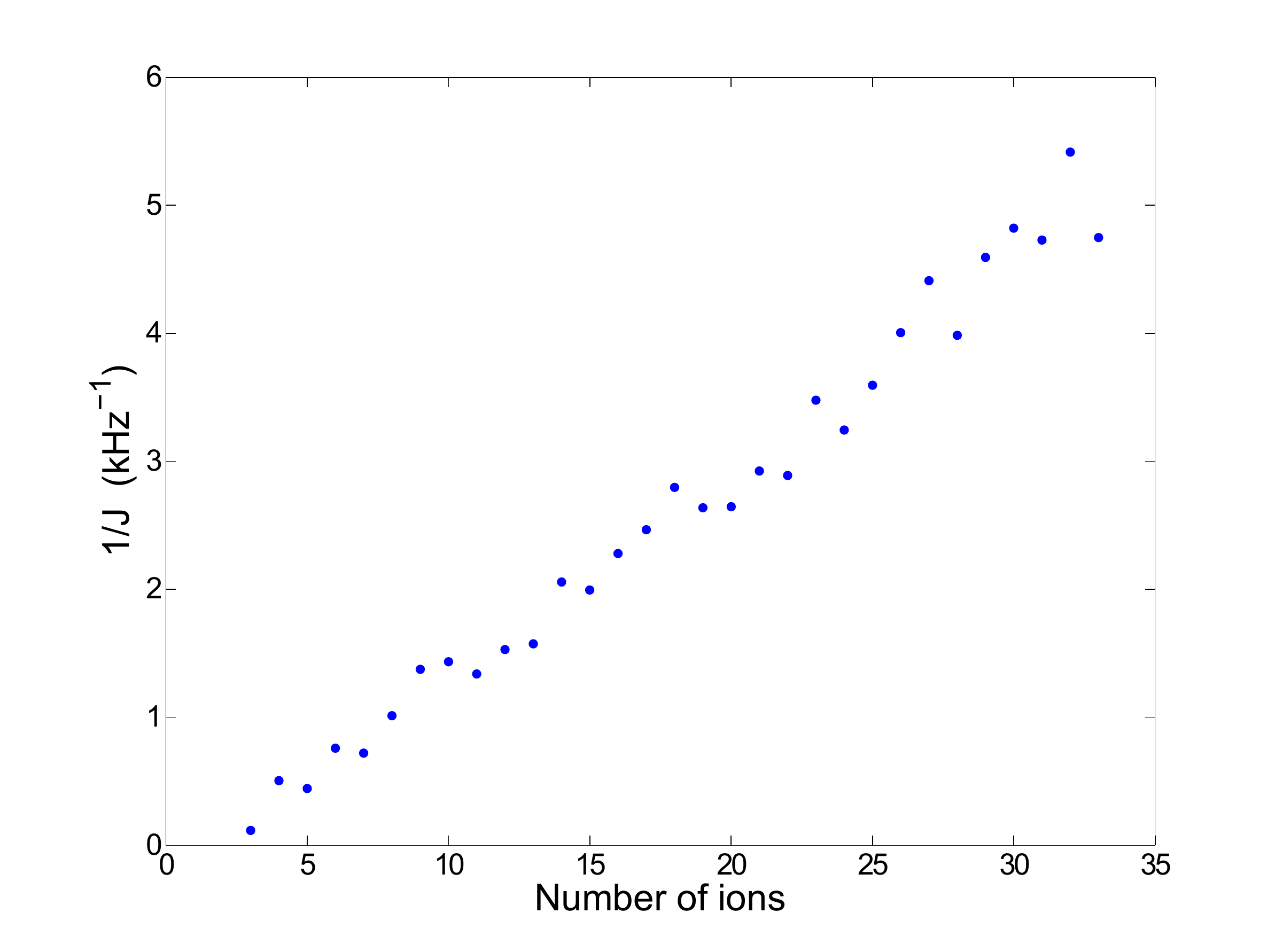}
\caption{Scaling of the nearest-neighbor coupling for the case of the production of a $1$D Ising chain 
for $N=3$ to $N=33$ ions. The total optical intensity is fixed with $\sum_{i,n} |\Omega_{i,n}| = 1$ \MHz, and each spectral component is detuned from its motional sideband by a fraction $f_s=0.03$ of smallest mode splitting, with a center-of-mass (COM) frequency $\omega_1/2\pi = 5$ \MHz. We find that the resulting Ising coupling scales roughly as $1/N$.}
\label{fig:scale}
\end{figure}

As the number of spins $N$ grows, the optical modulation scheme becomes more complex, with either $N$ or $N^2$ elements required.  We now estimate how the Ising couplings are expected to scale with the number of spins along with 
errors due to experimental fluctuations, phonon creation and spontaneous emission scattering, assuming a fixed transverse mode bandwidth.
The probability of phonon creation scales as 
$p_{ph} = \sum_{i,m}\left( \frac{\eta_{i,m} \Omega_{i,m}}{\omega_m-\mu_m} \right)^2$.
The off-resonant optical dipole forces are accompanied by a finite rate of spontaneous emission scattering, given 
by $\Gamma = \epsilon \sum_{i,m} |\Omega_{i,m}|$, 
where $\epsilon \ll 1$ is the ratio of excited state linewidth to Raman detuning.
The scaling of these potential errors depends upon the particular graph, so we consider two extremes. 
A uniform fully-connected interaction graph can be trivially generated with a single spectral component tuned close to the COM mode with a detuning $|\omega_1-\mu|/\omega_1 \ll \text{log}N/N^2$.
For a fixed level of phonon error, the total optical intensity should be reduced as $\text{log}N/N$, 
taking into account the intensity reduction per ion as the beam is expanded to accommodate the linearly expanding 
chain in space.  In this case the uniform Ising coupling is expected to scale as 
$N|J_{i,j}| \propto \text{log}N/N^2$, and the spontaneous emission rate per spin actually decreases with $N$.
For a sparse interaction graph, such as a $1$D (nearest-neighbor) Ising model, all modes are involved, and this time 
for a fixed phonon error the total optical intensity can remain fixed, since the typical
mode splitting falls only as $1/N$, while spontaneous emission per ion is fixed.  The calculation in Fig. \ref{fig:scale} shows that the resulting 
nearest-neighbor interaction scales as $J_{i,i+1} \propto 1/N$.  In either case of fully-connected or local Ising 
model, we thus expect to be able to support significant Ising interaction strengths with up to a few hundred spins.

For a general Ising graph, from Eq. \ref{J} we find that each pairwise interaction $J_{i,j}$ depends upon a balance of $N$ terms, and errors will accumulate with $N$ from fluctuations of relative optical intensities of the various spectral components of the beam (which should be stable if the spectral components are generated with high quality radiofrequency sources and modulators as shown in Fig. \ref{fig:schematic}) or their detunings from the motional sidebands.  The most important source will likely be fluctuations in the motional trap frequencies, where we expect the fractional error in the Ising coupling to grow as $\sqrt{N}(\delta \omega_m/\omega_m)$, so that a typical fractional fluctuation in the motional trapping frequencies of $\sim 10^{-3}$ might be expected to cause Ising coupling errors at a level of about $1\%$ for $N \sim 100$ ions.  

The scheme presented here can also be applied to more general Heisenberg spin models involving other noncommuting 
spin-spin interactions, such as the $XY$ model or the $2$D hexagonal Kitaev model relevant to topological 
quantum degrees of freedom \cite{Kitaev06}.  Here, additional Raman beams that couple to the other axes of motion can 
be exploited \cite{Porras04}.  Alternatively, a single direction of motion can be used as discussed above, with a stroboscopic 
alternation between Raman laser beams with different beat note phases as the various Ising interactions in the Hamiltonian 
are applied sequentially.  Here, we employ the Trotter expansion of the evolution operator \cite{Lloyd96, Lanyon11} and switch 
the various Ising terms rapidly enough so that higher order terms in the expansion can safely be neglected.
Although this discussion concentrated on a linear array of ions in space, these ideas apply in general to any stable ion crystal where the motional sidebands are resolved and prepared in the Lamb-Dicke limit, and should be useful for higher-dimensional trap geometries such as trap arrays \cite{Clark09, DidiSim09} or Penning traps \cite{Taylor08,Bollinger_2009}.

\begin{acknowledgments}
This work is supported by the US Army Research Office (ARO) with funds from the DARPA Optical Lattice Emulator (OLE) Program and the IARPA MQCO Program, the NSF Physics at the Information Frontier Program, the NSF Physics Frontier Center at JQI, and the National Basic Research Program of China (973 Program) 2011CBA00300 (2011CBA00302).

\end{acknowledgments}

\bibliography{arbitrary_simulation}

\end{document}